\documentclass[a4paper]{jpconf}
\usepackage{graphicx}
\begin{document}
\title{Magnetised Neutron Star Crusts and Torsional Shear Modes of Magnetars}

\author{Rana Nandi and Debades Bandyopadhyay}

\address{Astroparticle Physics and Cosmology Division,
Saha Institute of Nuclear Physics, 1/AF Bidhannagar,
Kolkata-700064, India}

\ead{debades.bandyopadhyay@saha.ac.in}

\begin{abstract}
We discuss outer and inner crusts of neutron stars in strong magnetic 
fields. Here, we demonstrate the effect of Landau quantization of electrons on 
the ground state properties of matter in outer and inner crusts in 
magnetars. This effect leads to the enhancement of the electron number density 
in strong magnetic fields with respect to the zero field case. For the outer 
crust, we adopt the magnetic Baym-Pethick-Sutherland model and obtain the 
sequence of nuclei and equation of state (EoS). The 
properties of nuclei in the inner crust in the presence of strong magnetic 
fields are investigated using the Thomas-Fermi model. The coexistence of two 
phases of nuclear matter - liquid and gas, is assumed in this case. The proton 
number density in the Wigner-Seitz cell is affected in strong magnetic fields 
through the charge neutrality. We perform this calculation using the Skyrme
nucleon-nucleon interaction with different parameterisations. We find nuclei 
with larger 
mass and atomic numbers in the inner crust in the presence of strong magnetic 
fields than those of the zero field case for all those parameter sets. 
Further we investigate torsional shear mode frequencies using the results of 
magnetised neutron star crusts and compare those with observations.  
\end{abstract}
\section{Introduction}
Neutron star crust plays an important role in many observational findings on 
neutron stars \cite{yak}. Heat transport and magnetic field evolution in the 
crust are 
sensitive to the composition of the crust. This, in turn, might influence 
various transport properties such as electrical and thermal conductivities,
shear modulus and 
shear viscosity in the neutron star crust. Figure 1 demonstrates different
regions of a neutron star schematically. The crustal region is just beneath the
envelope as shown in the figure. Neutron star crust is separated into outer and
inner crust. The outer crust begins at $\sim 10^4 g/cm^3$. The thickness of the
outer crust is a few hundred meters. The outer crust is made of nuclei arranged
in a body-centered cubic (bcc) lattice and embedded in a uniform background of
electrons which become relativistic at a density $\sim 10^{7} g/cm^3$. Neutrons and protons are bound in nuclei in this case. The
equilibrium nucleus at $\sim 10^4 g/cm^3$ is $^{56}$Fe. Equilibrium nuclei
become more and more neutron rich as density increases. The neutron chemical 
potential exceeds the neutron rest mass at a density 
$\sim 4 \times 10^{11} g/cm^3$. 
Consequently, neutrons drip out of nuclei and it is called the neutron drip 
point. This is the end of the outer crust and the beginning of the inner crust.
The inner crust is composed of neutron rich nuclei arranged in a bcc lattice
and also in coexistence with dripped
neutrons and a uniform background of electrons. The thickness of the inner
crust is $\sim$ 1 km. More and more neutrons come out of neutron rich
nuclei with increasing density. Finally, nuclei of the inner crust dissolve 
into a uniform nuclear matter at the crust-core boundary.       
\begin{figure}[h]
\begin{center}
\includegraphics[width=9cm]{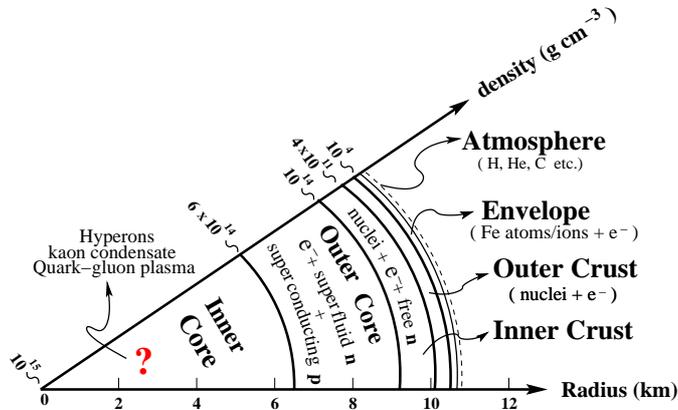}\hspace{2pc}%
\end{center}
\caption{\label{fig1}Schematic structure of a neutron star interior.}
\end{figure}

Extreme physical condition such as very strong surface magnetic fields 
$\sim 10^{15}$ G were found to exist in a class of neutron stars called 
magnetars \cite{dun92,dun98}. Even stronger interior field could exist in 
magnetars. Magnetars are classified into two groups - Anomalous x-ray pulsars 
(AXPs) and soft gamma-ray repeaters (SGRs) \cite{kou98,kou99}. SGRs are found
to emit sporadic bursts of soft gamma rays. It was observed that
SGRs occasionally emit stronger gamma rays known as giant flares having
luminosities as high as $10^{46}$ ergs s$^{-1}$ \cite{bar,isr,watt1,watt2}. 
Giant flairs are considered to be the result of evolving magnetic field and its
stress on the magnetar crust. In the decaying tail of giant flairs, 
quasi-periodic oscillations (QPOs) were observed over long duration 
\cite{isr,watt1,watt2}. QPOs were identified as torsional shear modes of
magnetar crust \cite{dun98}. The study of QPOs might shed light on the
ground state properties of magnetar crust.   

Recent studies showed that high magnetic fields associated with magnetars would
influence the composition and EoS of neutron star crusts which, in turn, might
have impact on the QPO frequencies. In this paper, we study and discuss this
problem. The paper is organised in the following way. We discuss the composition
and EoS of neutron star crusts with or without magnetic field in section 2.
Torsional shear modes are discussed in section 3. We summarise in section 4. 

\section{Composition and EoS of Neutron Star Crust}
In a seminal paper, Baym, Pethick and Sutherland (BPS) studied the equilibrium
composition and EoS of the ground state matter of the outer crust without
magnetic field \cite{bps}. Later this BPS model was extended to include the
effects of strong magnetic fields \cite{shapiro}. High magnetic fields of 
magnetars might influence the composition and EoS through Landau
quantisation. Recently we have revisited 
the magnetic BPS model. Furthermore, we include the finite size effect in the 
lattice energy and adopted recent experimental and theoretical mass tables 
\cite{rana1}. We adopt the Wigner-Seitz approximation in this calculation.
The equilibrium sequence of nuclei is determined by minimising 
the Gibbs free energy per baryon ($g$) with respect to mass number ($A$) and
atomic number ($Z$) at a constant pressure (P) \cite{bps},   
\begin{eqnarray}
g &=& \frac{E_{tot} + P}{n_b} = \frac{W_N + 4/3 W_L + Z \mu_e}{A}~,\nonumber\\
E_{tot} &=& n_N (W_N+ W_L)+ \varepsilon_e(n_e)~,\nonumber\\
P&=&P_e+\frac{1}{3}W_Ln_N~,
\end{eqnarray}
where $n_b$ is the baryon number density and $n_N$ is the number 
density of nuclei; those two are connected by $n_b = A n_N$.
The energy of the nucleus (including rest mass energy of nucleons) is
\begin{equation}
W_N = m_n (A - Z) + m_p Z - bA~,
\end{equation}
where $b$ is the binding energy per nucleon.
We obtain experimental nuclear masses from the atomic mass table 
of Ref.(\cite{audi03}). We exploit the theoretical extrapolation 
for the rest of nuclei \cite{moller95}. 
The lattice energy of the cell $W_L$ is given below \cite{bbp},
\begin{equation}
  W_L= -\frac{9}{10}\frac{Z^2e^2}{r_C}\left(1-\frac{5}{9} 
\left(\frac{r_N}{r_C}\right)^2\right)~.
\label{lat}
\end{equation}
The cell and nucleus radii are $r_C$ and $r_N$, respectively. The first term
in Eq.(\ref{lat}) is the lattice energy for point nuclei and the second term
corresponds to the finite size effect \cite{bbp}.

Now we focus on the impact of strong magnetic fields on the Gibbs free energy 
per baryon. In presence of a magnetic field, the motion of charged particles
perpendicular to the magnetic field is quantised into Landau levels. This is
known as Landau Quantisation \cite{lai01}. In this calculation, we are 
interested only in Landau quantisation of electrons and its influence on ground 
state properties of neutron star crusts. 
The electron energy density ($\varepsilon_e$), pressure ($P_e$) and
number density ($n_e$) due to Landau quantisation of electrons are given 
below \cite{rana1},
\begin{eqnarray}
\varepsilon_e &=&
\frac{eB}{4\pi^2} \sum_{\nu=0}^{\nu_{max}}g_{\nu} 
\left(p_{F_{e,\nu}}\mu_e + (m_e^2 + 2 e B \nu) 
\ln \frac{p_{F_{e,\nu}}+\mu_e}{\sqrt{(m_e^2 + 2 e B \nu)}}\right)~,\nonumber\\
P_e &=&
\frac{eB}{4\pi^2} \sum_{\nu=0}^{\nu_{max}}g_{\nu} 
\left(p_{F_{e,\nu}}\mu_e - (m_e^2 + 2 e B \nu) 
\ln \frac{p_{F_{e,\nu}}+\mu_e}{\sqrt{(m_e^2 + 2 e B \nu)}}\right)~,\nonumber\\
n_e&=&\frac{eB}{2\pi^2}\sum_{\nu=0}^{\nu_{max}}g_{\nu}p_{F_{e, \nu}}~,
\end{eqnarray}
where $B$ is the magnetic field, $\nu_{max}$ is the maximum Landau quantum
number, $p_{F_{e, \nu}}=\sqrt{\mu_e^2 - (m_e^2 + 2 \nu e B)}$, $\mu_e$ is the
electron chemical potential, and the spin degeneracy $g_{\nu} =1$ for $\nu =0$ 
and $g_{\nu}=2$ for other values of Landau quantum numbers. We define 
$B_{*}=B/B_c$, where the critical field for electrons is 
$B_c = 4.414 \times 10^{13}$ G. In the following paragraphs, we use magnetic
field strengths in terms of $B_{*}$.  

Electron number density is greatly influenced due to Landau quantization in 
strongly quantising magnetic fields. For $B_{*}\leq 10^{3}$, electrons populate
large number of Landau levels. Consequently, there is no change in electron
number density compared with the zero field situation. However, electron number
density strongly enhanced when the field is $B \geq 10^{17}$ G. This is 
attributed to the fact that electrons populate only the zeroth Landau level
\cite{rana1}. This might impact the composition and EoS of the outer crust.    
\begin{table}
\caption{\label{tab1} Sequence of nuclei in the outer crust in absence of 
magnetic fields. Elements with their charges ($Z$) and neutron numbers ($N$), 
maximum mass density ($\rho_{max}$), electron number density ($n_e$), electron 
chemical potential ($\mu$) and Gibbs free energy per nucleon ($g$) for $B=0$ 
are recorded here.}
\begin{center}
\begin{tabular}{|c|c|c|c|c|c|c|}
\hline
\ element & $Z$ & $N$ & $\hspace{0.5cm} \rho _{max}$ &$\hspace{0.2cm} n_e $ & $\mu_e  $ & g  \\
   &&& \hspace{0.2cm} \footnotesize {(g/cm$^3$) }&\footnotesize{(cm$^{-3}$)}& \footnotesize {(MeV)}&\footnotesize {(MeV)} \\
  \hline
  $^{56}$Fe& 26 & 30 & $8.01\times 10^{6}$ & $2.24\times 10^{30}$ & $\hspace{0.2cm}0.95$ & $930.60$ \\

  $^{62}$Ni& 28 & 34 & $2.71\times 10^{8}$ & $7.38\times 10^{31}$ & $\hspace{0.2cm}2.61$ & $931.32$ \\

  $^{64}$Ni& 28 & 34 & $1.33\times 10^{9}$ & $3.51\times 10^{32}$ & $\hspace{0.2cm}4.34$ & $932.04$ \\

  $^{66}$Ni& 28 & 34 & $1.50\times 10^{9}$ & $3.82\times 10^{32}$ & $\hspace{0.2cm}4.46$ & $932.09$ \\

  $^{86}$Kr& 36 & 50 & $3.10\times 10^{9}$ & $7.80\times 10^{32}$ & $\hspace{0.2cm}5.64$ & $932.56$ \\
  
  $^{84}$Se& 34 & 50 & $1.06\times 10^{10}$ & $2.58\times 10^{33}$ & $8.39$ & $933.62$ \\

  $^{82}$Ge& 32 & 50 & $2.79\times 10^{10}$ & $6.54\times 10^{33}$ & $11.43$ & $934.75$ \\

  $^{80}$Zn& 30 & 50 & $6.11\times 10^{10}$ & $1.37\times 10^{34}$ & $14.63$ & $935.90$ \\

  $^{78}$Ni& 28 & 50 & $9.25\times 10^{10}$ & $1.99\times 10^{34}$ & $16.56$ & $936.57$ \\
 
  $^{126}$Ru& 44 & 82 & $1.29\times 10^{11}$ & $2.69\times 10^{34}$ &$18.30$ & $937.12$ \\
 
  $^{124}$Mo& 42 & 82 & $1.86\times 10^{11}$ & $3.78\times 10^{34}$ & $20.50$ & $937.83$ \\
 
  $^{122}$Zr& 40 & 82 & $2.64\times 10^{11}$ & $5.18\times 10^{34}$ &$22.76$ & $938.53$ \\
 
  $^{120}$Sr& 38 & 82 & $3.77\times 10^{11}$ & $7.13\times 10^{34}$& $25.33$ & $939.31$ \\
 
  $^{118}$Kr& 36 & 82 & $4.34\times 10^{11}$ & $7.91\times 10^{34}$& $26.22$ & $939.57$ \\
 \hline
 \end{tabular}  
 \end{center}
 \end{table}  

We calculate the sequence of nuclei in the outer crust for $B=0$ and 
$B_{*}=10^{3}$. Results are shown in Table 1 and Table 2. A careful examination
of results in both tables shows that nucleus such as $^{78}$Ni which is
present for $B=0$, is absent in the case of $B_{*}=10^3$. Similarly, two new
nuclei $^{88}$Sr and $^{128}$Pd are found to appear in Table 2. The finite size
correction to the lattice energy modifies the sequence of nuclei for 
$B_{*}=10^4$. It is to be noted that our results are different from those of
Ref.\cite{shapiro} because we have used the recent experimental and 
theoretical nuclear mass tables in our calculation. One important result of 
this calculation is the shift of the neutron drip point to higher density in 
presence of strong magnetic fields. This is evident when we compare 
$\rho_{max}$ which is the maximum density for the existence of an equilibrium
nucleus, in the last row of Table 1 and Table 2. 
\begin{table}
\caption{\label{tab2} Same as Table 1 but for $B_{*} = 10^3$.}
  \begin{center}
  \begin{tabular}{|c|c|c|c|c|c|c|}
  \hline
   \ element & $Z$ & $N$ & $\hspace{0.5cm} \rho _{max}$ &$\hspace{0.2cm} n_e $ & $\mu_e  $ & g  \\
   &&& \hspace{0.2cm} \footnotesize {(g/cm$^3$) }&\footnotesize{(cm$^{-3}$)}& \footnotesize {(MeV)}&\footnotesize {(MeV)} \\
  \hline
  $^{56}$Fe& 26 & 30 & $4.31\times 10^{9}$ & $1.21\times 10^{33}$ & $\hspace{0.2cm}0.87$ & $930.46$ \\

  $^{62}$Ni& 28 & 34 & $1.83\times 10^{10}$ & $4.99\times 10^{33}$ & $\hspace{0.2cm}2.94$ & $931.32$ \\

  $^{64}$Ni& 28 & 34 & $2.33\times 10^{10}$ & $6.14\times 10^{33}$ & $\hspace{0.2cm}3.60$ & $931.59$ \\

  $^{88}$Sr& 38 & 50 & $2.59\times 10^{10}$  & $6.73\times 10^{34}$ & $\hspace{0.2cm}3.94$ & $931.72$ \\

  $^{86}$Kr& 36 & 50 & $4.33\times 10^{10}$ & $1.09\times 10^{34}$ & $\hspace{0.2cm}6.35$ & $932.69$ \\
  
  $^{84}$Se& 34 & 50 & $6.33\times 10^{10}$ & $1.54\times 10^{34}$ & $8.97$ & $933.72$ \\

  $^{82}$Ge& 32 & 50 & $8.67\times 10^{10}$ & $2.04\times 10^{34}$ & $11.83$ & $934.82$ \\

  $^{80}$Zn& 30 & 50 & $1.13\times 10^{11}$ & $2.55\times 10^{34}$ & $14.82$ & $935.93$ \\
 
  $^{128}$Pd& 46 & 82 & $1.29\times 10^{11}$ & $2.78\times 10^{34}$ & $16.15$ & $936.38$ \\
 
  $^{126}$Ru& 44 & 82 & $1.50\times 10^{11}$ & $3.15\times 10^{34}$ &$18.32$ & $937.13$ \\
 
  $^{124}$Mo& 42 & 82 & $1.72\times 10^{11}$ & $3.50\times 10^{34}$ & $20.35$ & $937.82$ \\
 
  $^{122}$Zr& 40 & 82 & $1.96\times 10^{11}$ & $3.86\times 10^{34}$ &$22.45$ & $938.50$ \\
 
  $^{120}$Sr& 38 & 82 & $4.34\times 10^{11}$ & $8.22\times 10^{34}$& $25.44$ & $939.32$ \\
 
  $^{118}$Kr& 36 & 82 & $4.92\times 10^{11}$ & $8.98\times 10^{34}$& $26.27$ & $939.57$ \\
 \hline
 \end{tabular}  
 \end{center}
 \end{table}  

Now we discuss the inner crust. Nuclei of the inner crust are not only immersed
in a uniform background of electrons but also in coexistence with a neutron gas.
Besides charge neutrality, the matter of the inner crust is in 
$\beta$-equilibrium. The ground state properties of the inner crust in zero
magnetic field were studied by several groups \cite{lan,beth,bay,neg,che}. The
first calculation of the inner crust in presence of strong magnetic fields was
investigated in the Thomas-Fermi (TF) model at zero temperature \cite{rana2}.    
Nuclei of the inner crust are also arranged in a bcc lattice. The Wigner-Seitz
approximation is used in the calculation of the inner crust in presence of 
strong magnetic fields. Further, each lattice volume is replaced by a spherical
cell with one nucleus at the center. Each cell is charge neutral. There is no 
Coulomb interaction between 
cells. In this case, the denser phase in a nucleus is in
coexistence with the low density phase of neutron gas. This has to be  dealt 
with in a thermodynamically consistent way. The spherical cell in which protons
and neutrons reside, does not define a nucleus. We exploit the subtraction 
procedure of Bonche, Levit and Vautherin (BLV) \cite{bon} to remove the gas 
part from the nucleus plus gas in the cell. In the BLV scheme, the density
profiles of the gas part as well as nucleus plus gas are derived 
self-consistently in the TF formalism. Finally, the difference of two solutions
in BLV scheme determines the nucleus \cite{bon,sur}. Here electrons are Landau
quantised in strong magnetic fields, but protons in the cell are indirectly
affected through the charge neutrality.  

Our starting point in this calculation of inner crust is the thermodynamical 
potential ($\Omega_N$) of a nucleus. This is obtained as the difference of the 
thermodynamical potential in nucleus plus gas ($\Omega_{NG}$) and that of the
gas phase ($\Omega_G$). This is given by \cite{bon},
\begin{equation}
\Omega_N = \Omega_{NG} - \Omega_{G}~,
\end{equation}
where the thermodynamic potential is written in terms of free energy($F$), 
chemical potential ($\mu_i$) and number density ($n_i$) of i-th species, 
\begin{equation}
\Omega = F - \sum_{i=n,p} \mu_i n_i~.
\end{equation} 
The free energy is a function of average baryon density ($n_b$) and proton
fraction ($Y_p$) and given below,
\begin{equation}
F(n_b,Y_p) = \int [\varepsilon_N + \varepsilon_c + \varepsilon_e] d{\bf r}~.
\end{equation}
The free energy contains nuclear energy density ($\varepsilon_N$), Coulomb 
energy
density ($\varepsilon_c$) and electron energy density. We calculate the nuclear
energy density using Skyrme (SkM) nucleon-nucleon interaction 
\cite{rana2,kri,bra,sto}. We minimise the thermodynamical potential 
with respect to baryon density in each phase and obtain neutron and proton
density profiles of two phases solving coupled equations \cite{rana2,sil}. The
chemical potential in presence of magnetic fields is \cite{rana2}    
\begin{equation}
\mu_e = \left[p_{F_{e,\nu}}^2+ m_e^2 + 2 e B \nu \right]^{1/2} - <V^c(r)>~,
\end{equation}
where $<V^{c}(r)>$ denotes the average single particle Coulomb potential. 
As soon as we know density profiles in two phases, we calculate mass and atomic
numbers of a nucleus using the subtraction procedure of BLV \cite{rana2}. 

\begin{figure}[h]
\begin{center}
\includegraphics[width=9cm]{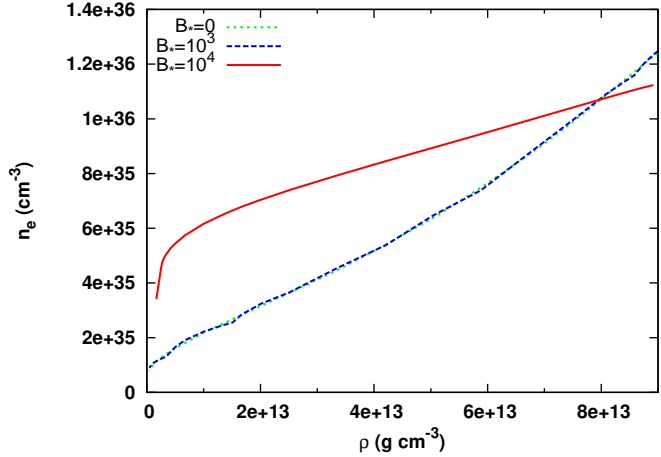}\hspace{2pc}%
\end{center}
\caption{\label{fig2}Electron number density is plotted with mass density for 
different magnetic field strengths.}
\end{figure}

Now we present our results of the inner crust. As electrons are Landau 
quantised in strong magnetic fields, we show its impact on electron number
density. Figure 2 shows electron number density as a function of mass density.
For $B_{*} = 10^3$ or smaller field strength, large numbers of Landau levels
are populated by electrons. As a result, the electron number density does not 
differ from the zero field results. For stronger magnetic fields such as
$B_{*} = 10^4$, electrons populate the zeroth Landau level or a few levels
over a certain mass density regime. In this case, it is evident from 
Fig. \ref{fig2}
that electron number density is significantly enhanced compared with the zero
field case. However, the situation is different at higher densities. We do not
see any enhancement of electron fraction there. Proton fraction is modified
in presence of strong magnetic fields through charge neutrality. Neutrons are
affected by magnetic fields through $\beta$-equilibrium. The outcome of this
is fewer neutrons drip out of a nucleus in presence of strong magnetic fields
\cite{rana2}.

\begin{figure}[th]
\centerline{\includegraphics[width=9cm]{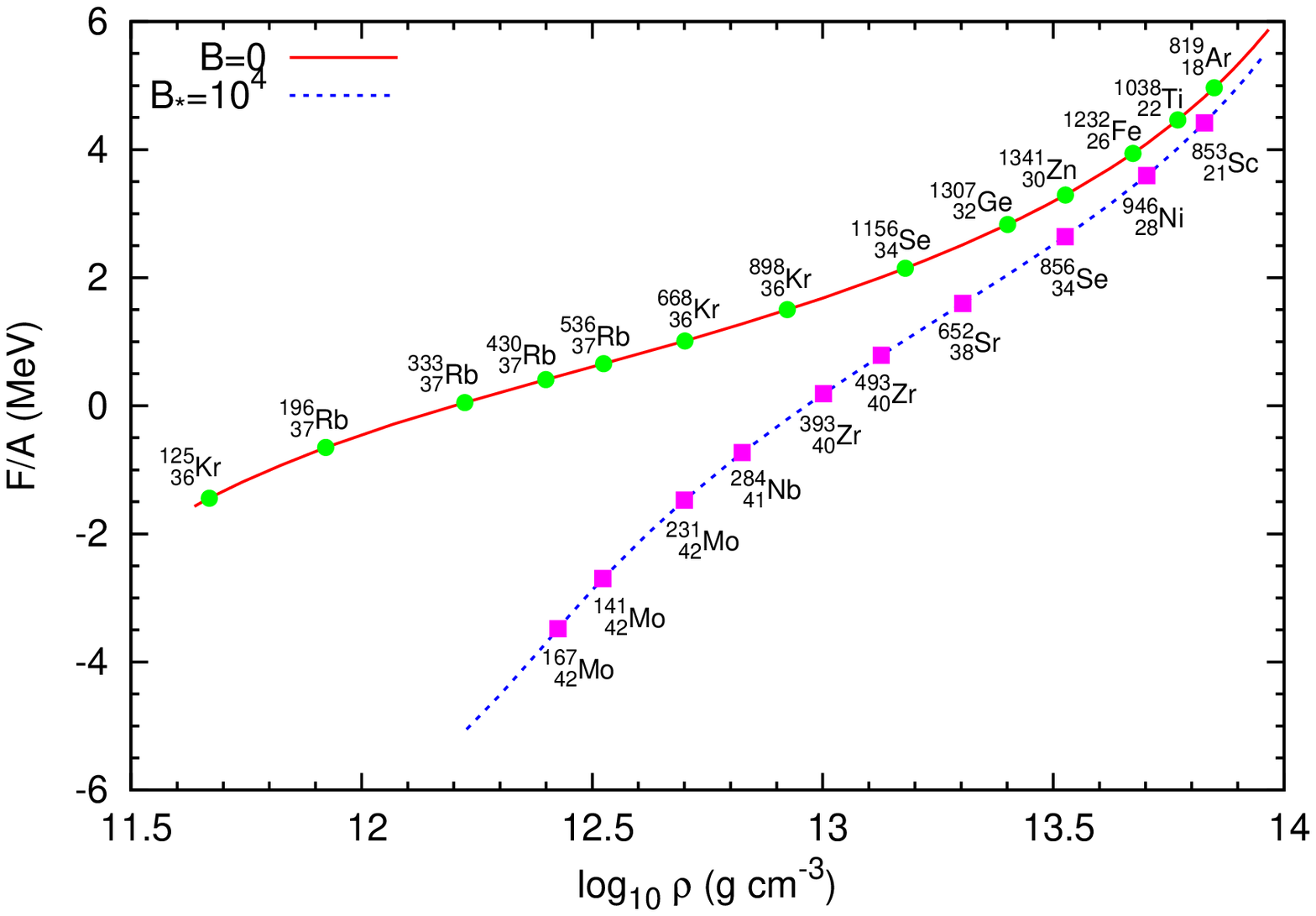}\includegraphics[width=9cm]{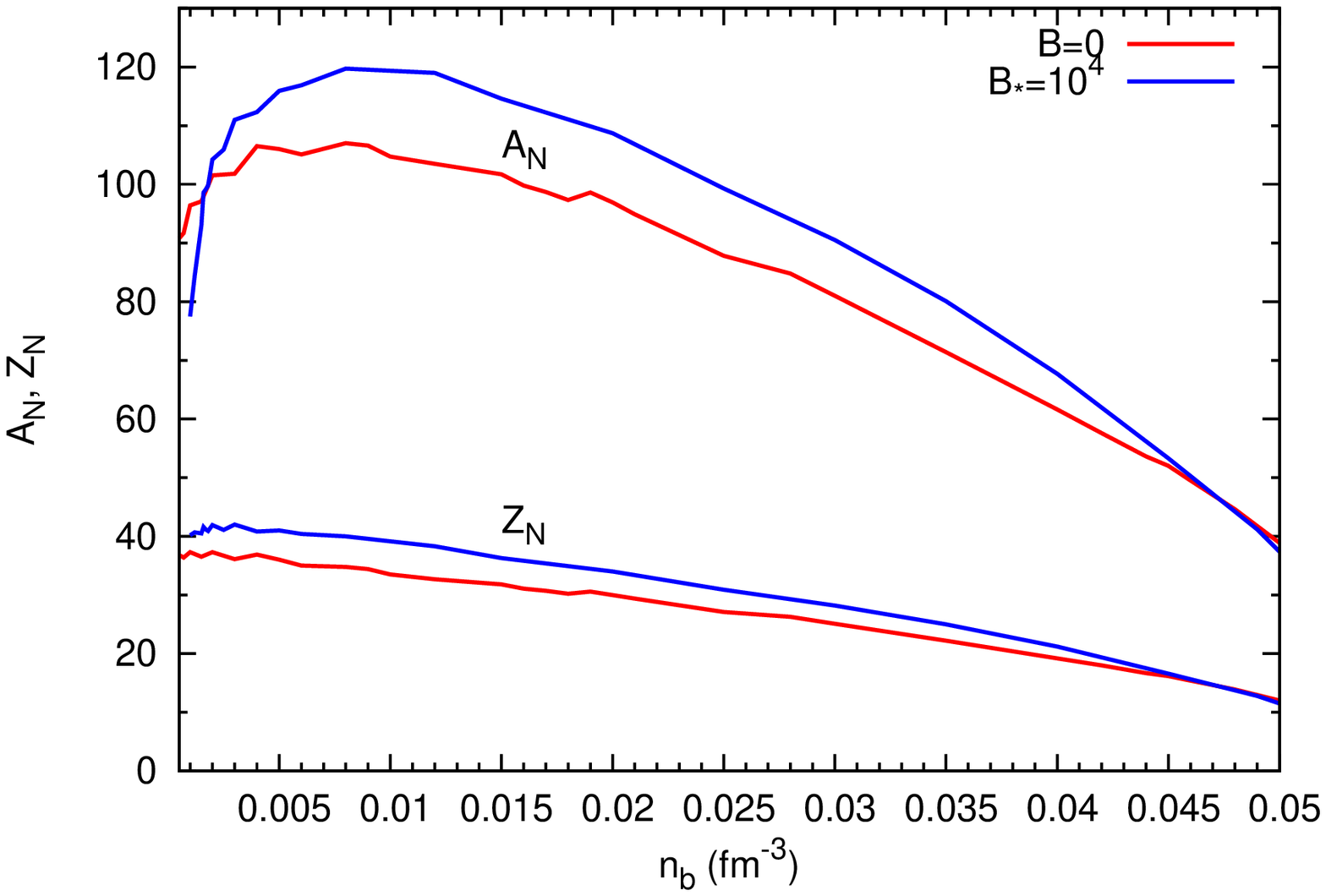}}
\caption{Free energy per nucleon of the cell is plotted with mass
density for zero magnetic field, $B=10^3$ and $B_{*}=10^4$ (left panel). 
Equilibrium nuclei are denoted with solid symbols in both panels; mass and 
atomic numbers in nuclei after the subtraction of the gas phase is shown as a
function of average baryon number density (left panel).}
\label{fig3}       
\end{figure}

Next we obtain the equilibrium nucleus at each density point by minimising the
free energy per particle of the cell \cite{rana2}. In the left panel of 
Fig. \ref{fig3}, we show the sequence of equilibrium nuclei and their 
corresponding free energies per particle as a function of mass
density with and without magnetic field. Furthermore, this calculation is
done without subtracting the gas part. It is noted that the free 
energy per nucleon is reduced in strong magnetic fields of $B_{*}=10^4$. 
This is attributed to the population of electrons in the zeroth Landau level.
However, free energy per nucleon for $B=0$ and $B_{*}=10^4$ come closer when
large number of Landau levels are populated at higher densities. 
For zero magnetic field case, we find excellent
qualitative agreement between our results and those of Negele and Vautherin
\cite{neg}. In the right
panel of Fig. \ref{fig3}, mass and atomic numbers of equilibrium nuclei are 
shown
after applying the subtraction method of BLV. We find that mass and atomic
numbers increase in presence of a strong magnetic field compared with the 
results of zero magnetic field. 

\begin{figure}[th]
\centerline{\includegraphics[width=9cm]{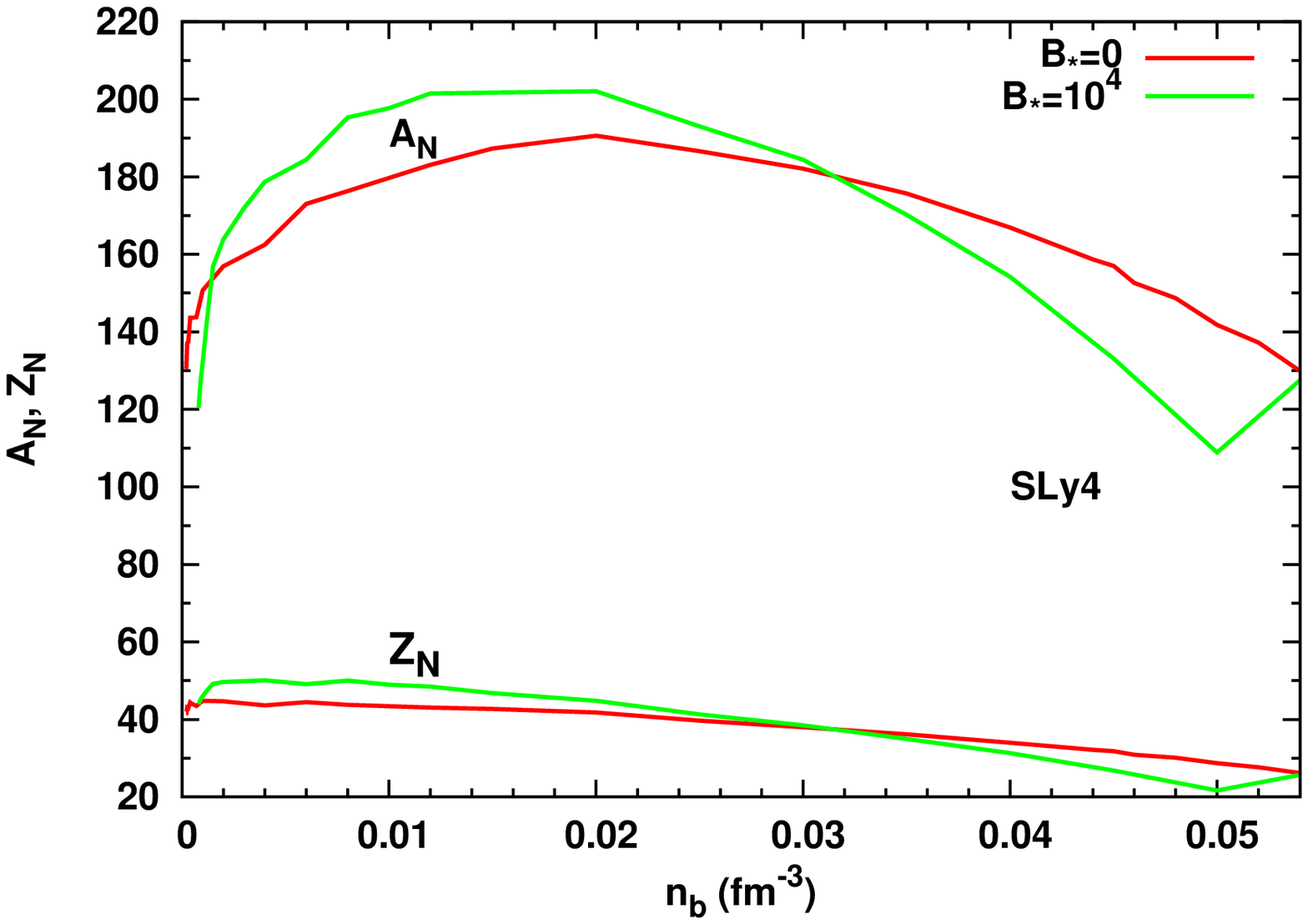}\includegraphics[width=9cm]{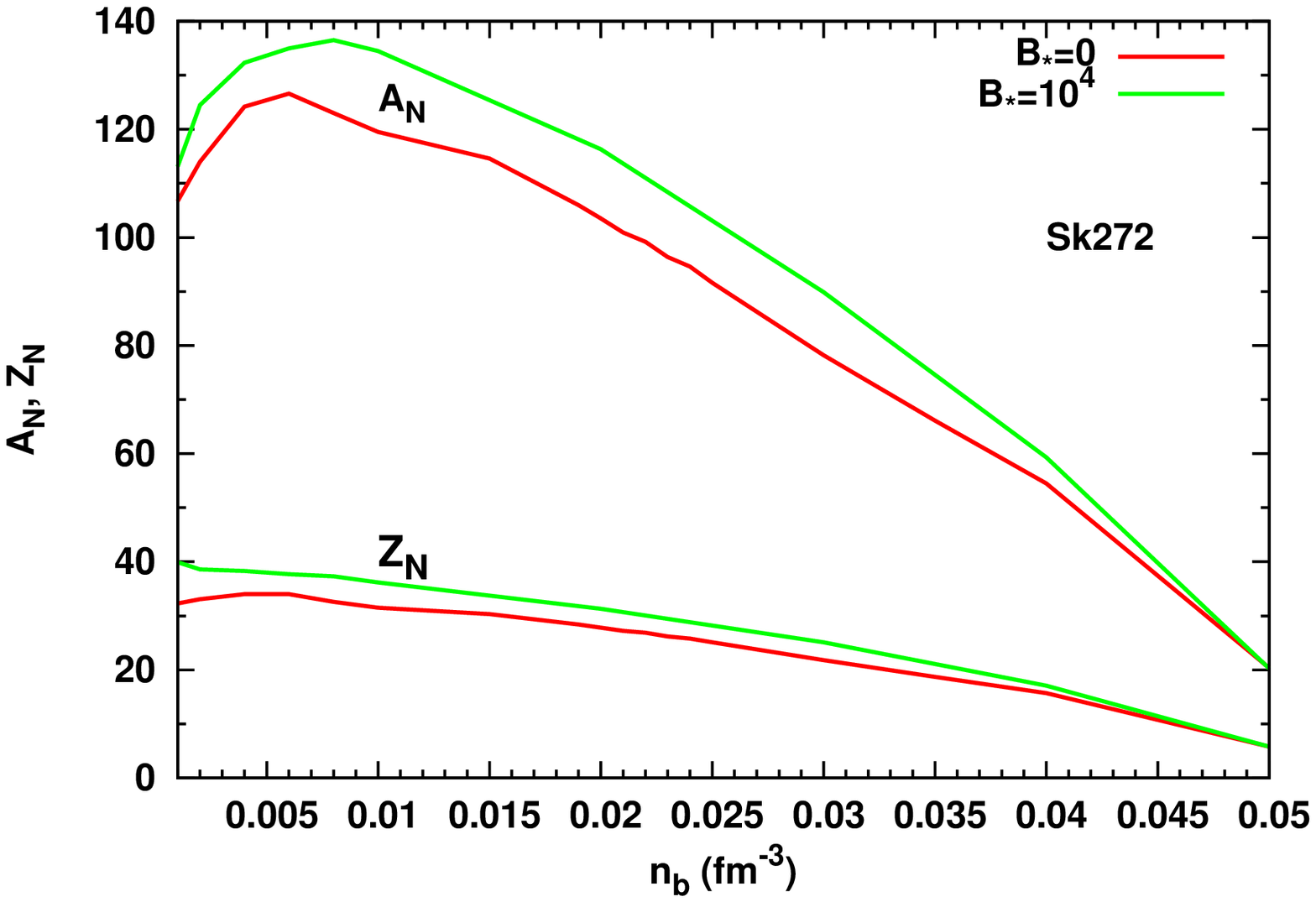}}
\caption{
Mass and 
atomic numbers in nuclei after the subtraction of the gas phase is shown as a
function of average baryon number density for SLy4 interaction (left panel) as 
well as Sk272 (right panel) with $B=0$ and $B_{*}=10^4$.}
\label{fig4}       
\end{figure}

We repeat the calculation of inner crust with SLy4 \cite{chaba} and Sk272 
\cite{bka} nucleon-nucleon interactions. Results of these calculations are 
shown
in Fig. \ref{fig4}. We find qualitatively similar results for mass and atomic 
numbers as a function of average baryon number density as found in case of SkM
interaction. For SLy4 interaction, we find mass and atomic numbers in the
magnetic field jump when electrons in the zeroth Landau level move to the next
level around baryon number density 0.05 $fm^{-3}$. Further, we note that the 
SLy4 parameter set results in higher 
mass and atomic numbers because of stiffer density dependence of the symmetry
energy at sub-saturation densities than that of other parameter sets considered
here. 

\section{Torsional Shear Modes}
Next we are interested in the role of compositions and EoS of magnetised crusts
that we have described in section 2, on torsional shear modes of magnetars. It 
is argued that torsional shear mode frequencies depends on the shear modulus of
neutron star crusts \cite{watt1,watt2,stei}. Further the shear modulus is
sensitive to compositions and EoS of neutron star crusts. This implies that 
nuclear physics plays an important role in determining torsional shear 
frequencies \cite{stei}. Here we describe the calculation of the shear modulus 
and shear velocity of the magnetised neutron star crusts. We use the following
the expression of the shear modulus($\mu$) at zero temperature
\cite{ichi,stroh},  
\begin{equation}
\mu = 0.1194 \frac{n_i (Ze)^2}{a}~,
\label{shr}
\end{equation}
where $a = 3/(4 \pi n_i)$, $Z$ is the atomic number of a nucleus and $n_i$ is
the ion density. This form of the shear modulus was 
obtained by assuming a bcc lattice and performing directional averages 
\cite{han}. We exploit inputs $n_i$ and $a$ from the magnetised crusts of 
section 2. We immediately calculate the shear speed as 
$v_s = \sqrt{(\mu/\rho)}$.
In Figure 5, the shear speed is shown as a function of mass density with and
without magnetic field. For $B_{*}=10^3$, the shear speed does not differ from
that of $B=0$. In presence of strong fields such as $B_{*}=10^4$, the shear 
speed enhanced compared with the zero field results. We find spikes in each 
curve. It happens when an equilibrium nucleus jumps to a new one in the outer
crust.  
\begin{figure}[h]
\begin{center}
\includegraphics[width=9cm]{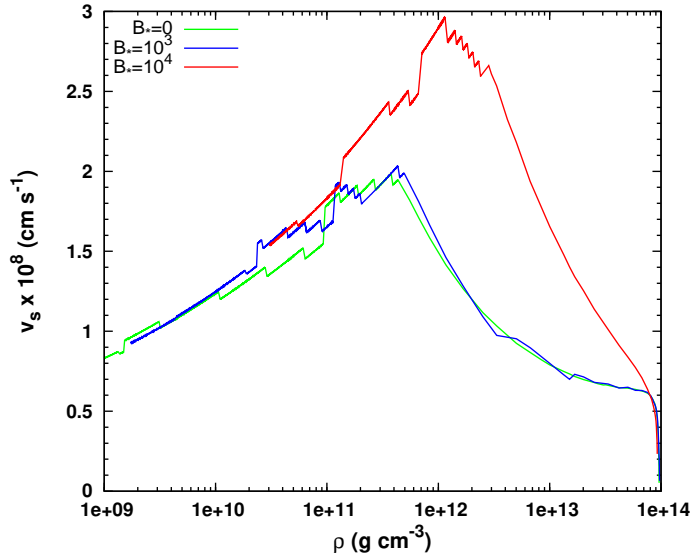}\hspace{2pc}%
\end{center}
\caption{Shear velocity is shown as a function of mass density for 
different magnetic field strengths.}
\label{fig5}       
\end{figure}

Torsional shear modes were studied extensively using non-magnetic crusts 
in Newtonian gravity \cite{dun98,piro,carr,mcd} and general relativity 
\cite{sota1,sota2,sota3,kip,mes}. Here we are interested in the effects of
magnetised crusts on torsional shear mode frequencies. We adopt the model of
Refs.\cite{sota1,mes} to calculate torsional shear frequencies in presence 
of a dipole magnetic field. 
In this calculation, we consider the magnetised crust decoupled from the core. 
Further, the magnetised star is considered to be the spherically symmetric
because the deformation due to the magnetic field is neglected \cite{mes}.
The perturbed equations describing torsional shear modes are obtained by 
linearising the equations that govern equilibrium configurations. Equilibrium
stellar models are determined using the metric
\begin{equation}
ds^2 = - e^{2\Phi} dt^2 + e^{2\Lambda} dr^2 + r^2 \left( d{\theta}^2 + sin^2{\theta} d{\phi}^2 \right)~.
\end{equation}
We consider axial-type perturbation in the four velocity ($u^{\mu}$) and the 
only
non-vanishing perturbed quantity is the $\phi$ component of the perturbed four
velocity $u^{\phi}$ \cite{sota1}
\begin{equation}
\partial {u^{\phi}} = e^{-\phi} \partial_t {\cal{Y}}(t,r) {\frac{1}{sin{\theta}}}{\partial_{\theta}} P_l(cos{\theta})~,
\end{equation}
where $\partial_t$ and $\partial_{\theta}$ correspond to partial derivatives
with respect to time and $\theta$, respectively, $P_l(cos{\theta})$ is 
the Legendre polynomial of order $\ell$ and ${\cal{Y}}(t,r)$ is the angular 
displacement of the matter. Assuming a harmonic time dependence of 
${\cal{Y}}(t,r)$, one arrives at the eigenvalue equation which is a second
order differential equation \cite{sota1}. Finally, we estimate eigenfrequencies
by solving two first order differential equations Eq.(69) and (70) of Sotani 
et al. with appropriate boundary conditions \cite{sota1}.

\begin{figure}[th]
\centerline{\includegraphics[width=8cm]{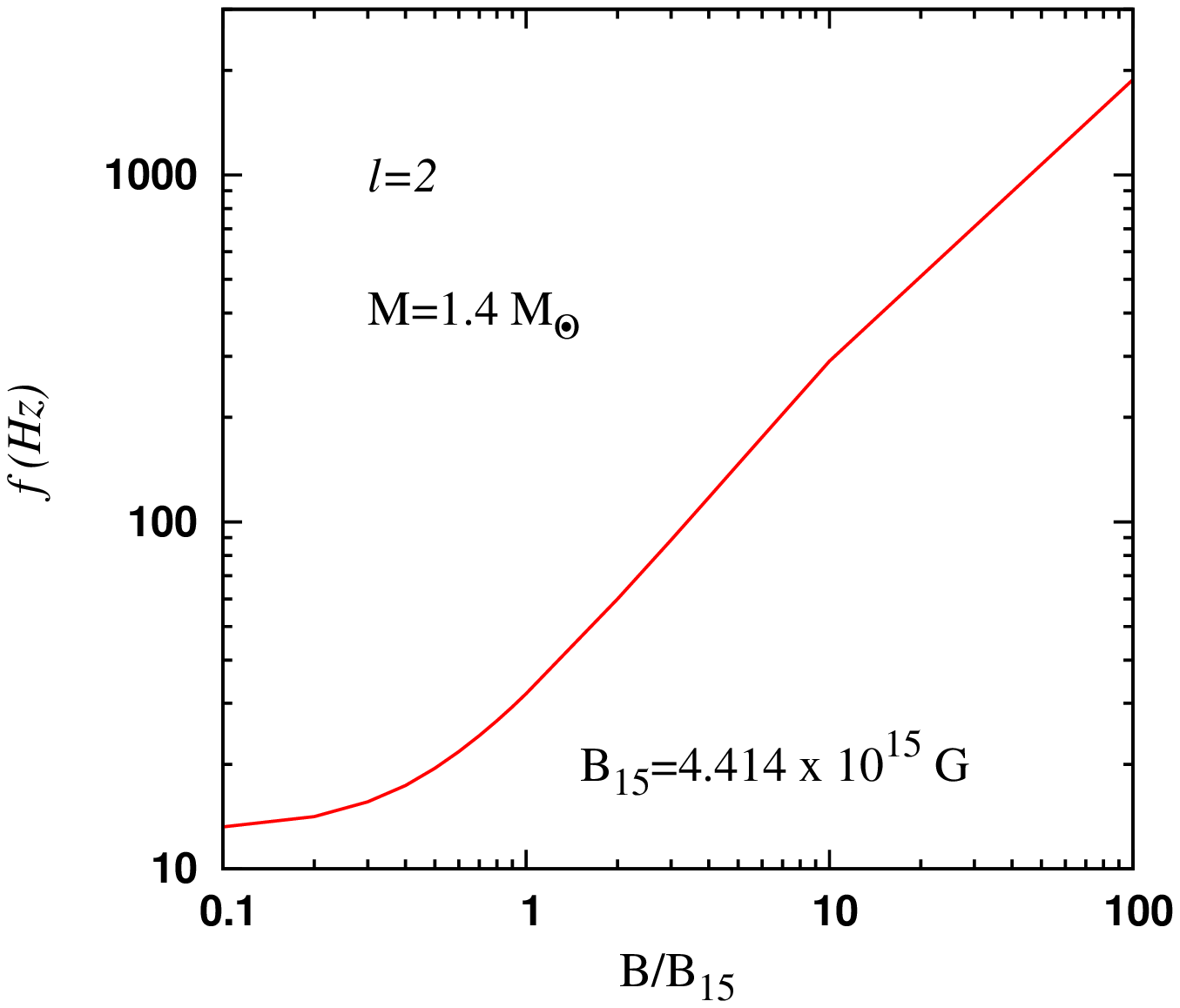}\includegraphics[width=10cm]{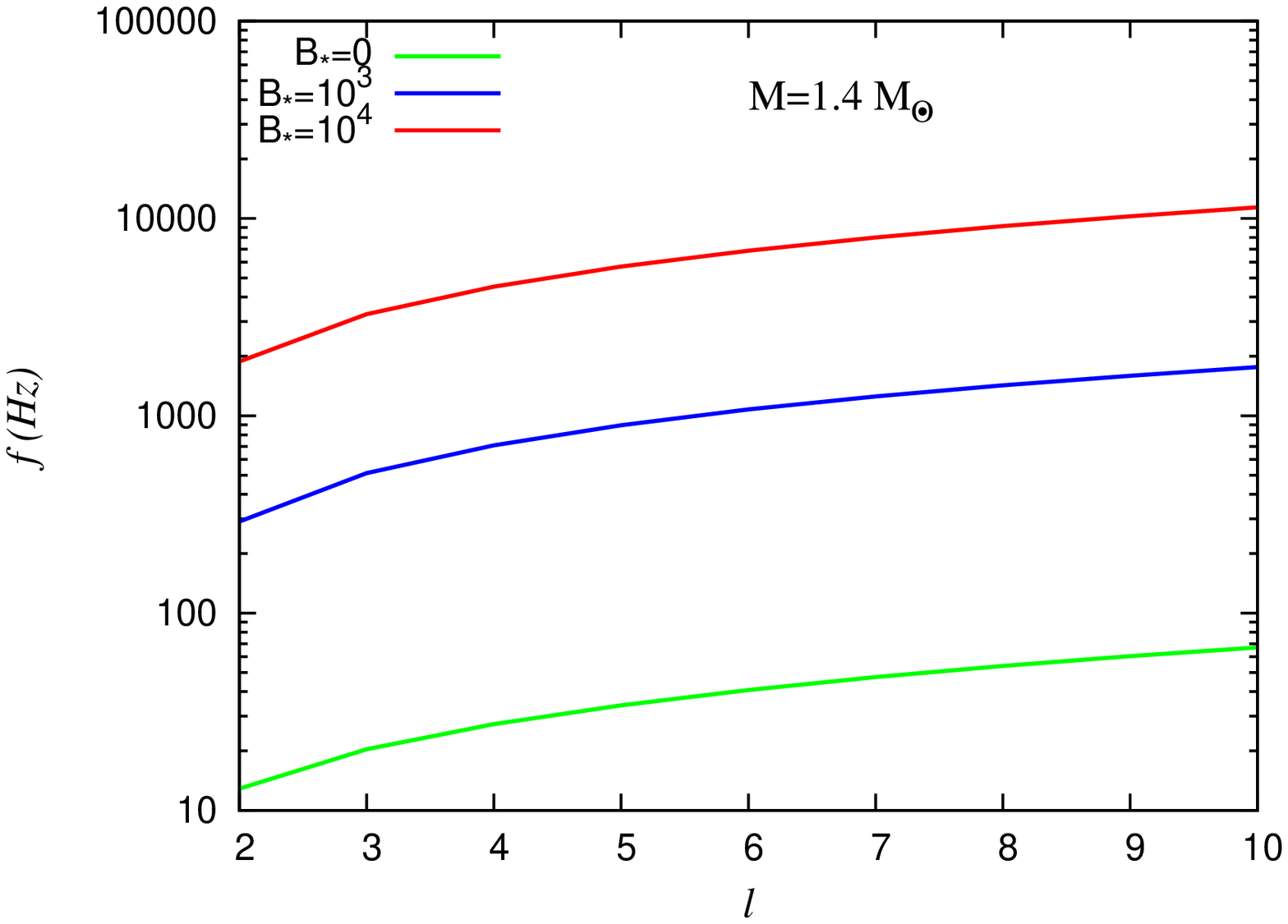}}
\caption{Fundamental ($n=0$, $\ell=2$) shear mode frequency as a function of 
normalised magnetic field for a neutron star of 1.4 $M_{\odot}$ 
(left panel); Torsional shear mode frequency ($n=0$) as a function of $\ell$ 
values for different magnetic field strengths and a neutron star of 1.4 
$M_{\odot}$ (right panel).}
\label{fig6}       
\end{figure}

Now we discuss the results of our calculation using the magnetised crusts
calculated with SkM nucleon-nucleon interaction. 
In
the left panel of Figure 6, fundamental shear mode frequency ($n=0$, $\ell=2$)
is plotted with magnetic field for a neutron star of 1.4 $M_{\odot}$. The 
fundamental frequency increases with increasing magnetic field. This plot 
implies if we know the mass of the star and the field strength accurately,
we can determine the frequency and compare it with the observed frequency.
In the right panel of Figure 6, we show torsional shear mode frequency for
$n=0$ as a function of $\ell$ values. As $\ell$ values increase, the frequency
increases. It is also evident from this figure that the torsional frequency
increases with increasing magnetic field strengths. It is also noted that
frequencies of torsional shear modes decrease with increasing masses of 
magnetars \cite{rana3}.

\begin{table}
\caption{\label{tab3} Calculated torsional shear mode frequencies are compared
with observed frequencies of SGR 1806-20.}
\begin{center}
 \begin{tabular}{|c|c|c|c|}
 \hline
 \multicolumn{4}{|c|}{SGR 1806-20}\\
 \cline{1-4}
 Observed & Calculated & & \\
 Frequency & Frequency & n & $\ell$\\
 (Hz) & (Hz) &&\\
 \hline\hline
 18 & 15 & 0 & 2\\
 26 & 24 & 0 & 3\\
 29 & 32 & 0 & 4\\
 93 & 93 & 0 &12\\
 150 & 151 & 0 & 20\\
 626 & 626 & 1 & 29\\
 1838 & 1834 & 4 & 2\\
 \hline
 \end{tabular}
\end{center}
\end{table}

\begin{table}
\caption{\label{tab4} Same as Table 3 but for SGR 1900+14.}
\begin{center}
 \begin{tabular}{|c|c|c|c|}
 \hline
 \multicolumn{4}{|c|}{SGR 1900+14}\\
 \cline{1-4}
 Observed & Calculated & & \\
 Frequency & Frequency & n & $\ell$\\
 (Hz) & (Hz) &&\\
 \hline\hline
 28 & 28 & 0 & 4\\
 54 & 55 & 0 & 8\\
 84 & 82 & 0 &12\\
 155 &154 & 0 &23\\
  \hline
 \end{tabular}
\end{center}
\end{table}
Finally, we calculate fundamental torsional shear mode frequencies as well as
various overtones for SGR 1806-20 and SGR 1900+14
and compare those with observed frequencies of QPOs. We show this comparison
in Table 3 and Table 4. In case of SGR 1806-20,
higher torsional frequencies i.e. (93, 150, 626 and 1838 Hz) are nicely 
explained by our theoretical calculation with magnetised crusts when the
magnetic field is $B = 8 \times 10^{14}$ G and mass is 1.4 $M_{\odot}$ 
\cite{rana3}. However,
the agreement of three lower frequencies (18, 26 and 29 Hz) with our 
calculation is not good. These frequencies could possibly be explained by the 
Alfv\'en
modes \cite{colai}. On the other hand, all four observed frequencies 
(28, 55, 82, 154 Hz) in SGR 1900+14 are in excellent agreement with our 
calculation when
the mass is 1.2 $M_{\odot}$ and magnetic field $B = 4 \times 10^{14}$ G
\cite{rana3}.  

\section{Summary}
We have developed a model of magnetised neutron star crusts. In this context,
we have investigated the effects of Landau quantisation of electrons on the
ground state properties of outer and inner crusts. It is observed that
composition and EoS of neutron star crusts are appreciably modified in strong
magnetic fields. We have calculated the shear modulus and shear velocity in
connection with torsional shear modes. It is found that torsional shear
frequencies calculated using our model of magnetised crusts are in good 
agreement with the observed frequencies of QPOs.

\end{document}